\documentclass{article}
\usepackage[T1]{fontenc}
\usepackage[utf8]{inputenc}
\usepackage{ismir} 
\usepackage{amsmath,cite,url}
\usepackage{graphicx}
\usepackage{subcaption}
\usepackage{color}
\usepackage{tikz}
\usepackage{booktabs}
\usetikzlibrary{arrows.meta, positioning, decorations.pathmorphing, fit, shapes.geometric, matrix}
\usepackage{xcolor}
\usepackage{multirow}

\title{ Explore This! \\ Beat and Downbeat Tracking From a Learned Tatum Grid}

\threeauthors
   {Robert Kihlborg} {KTH Stockholm \\ robertok@kth.se}
   {André Holzapfel} {KTH Stockholm \\ holzap@kth.se}
   {Jan Schlüter} {JKU Linz \\ jan.schlueter@jku.at}

\def\authorname{R. Kihlborg, A. Holzapfel and J. Schlüter}

\usepackage[bookmarks=false,pdfauthor={\authorname},pdfsubject={\pdfsubject},hidelinks]{hyperref}

\begin{document}

\maketitle

\begin{abstract}
State-of-the-art beat tracking is based on neural networks that model beats and downbeats as individual events, often followed by a separate Dynamic Bayesian Network (DBN) enforcing musical consistency. We propose a novel architecture that combines deep learning with differentiable theory-based modules: Specifically, we design a system that first predicts a tatum grid, then groups tatum events to beats and downbeats, explicitly modelling multiple options for the current tempo and time signature in a bar-pointer-like fashion. Our approach improves continuity metrics compared to existing DBN-free systems and is more versatile than the de-facto standard DBN, but still falls behind the best DBN-based methods on average. Given the many design choices and hyperparameters involved, we assume we have not yet explored the full potential of this idea and publish our code and invite others to explore this further.
\end{abstract}

\section{Introduction}\label{sec:introduction}
Beat/downbeat tracking is the task of estimating the temporal positions of beats/downbeats in a music recording.
In the past decade, deep learning approaches have come to dominate this task, driven by improvements in architectures, as well as the increased availability of data and compute, allowing for cheaper and faster training. Most systems follow some variant of the same strategy: (1) The audio is converted to a spectrogram and fed to a neural network that outputs framewise beat/downbeat activations. (2) The network is trained with a loss that pushes the activation to be high at beats/downbeats and low otherwise, so it can be interpreted as a local probability of beat/downbeat appearance. (3) For inference, activations are postprocessed to materialize beats and downbeats, either by using a Dynamic Bayesian Network (DBN) or by finding large local maxima of the activation function.

\begin{figure}
    \centering
    \includegraphics[width=0.9\linewidth]{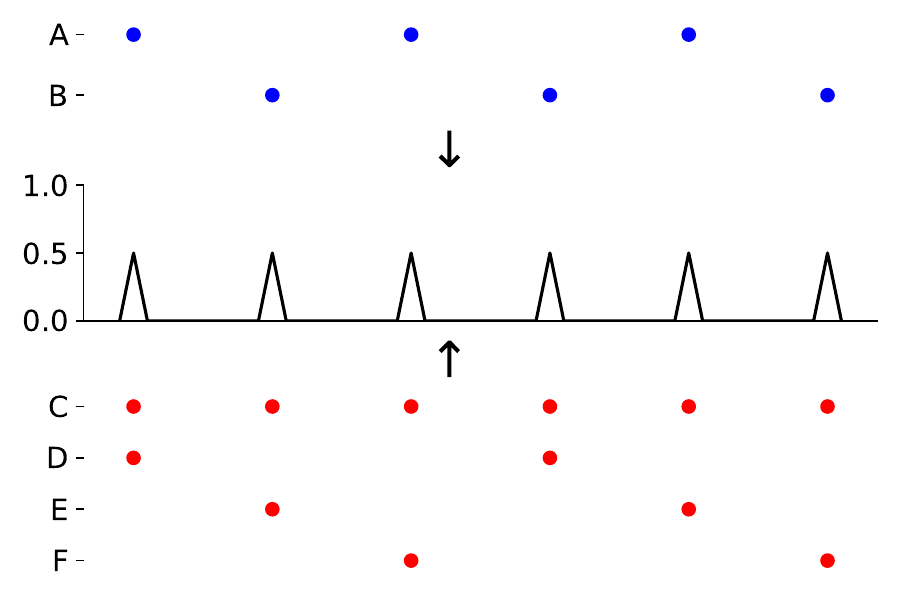}
    \caption{Illustration of how independent framewise probabilities can never encode ambiguity. A sample that is ambiguous between A and B (top) will produce the same local probabilities (middle) as a sample that is ambiguous between C, D, E and F (bottom).}
    \label{fig:rhythmic_ambiguity}
\end{figure}

Though it performs well in general, this strategy has foundational shortcomings. The activation function models probabilities independently, and therefore cannot express interdependence between probabilities at the various metrical levels. This means that it cannot handle metric ambiguity -- an ambiguous rhythm will lead to a mixture distribution that cannot be traced back to its constituents (Figure~\ref{fig:rhythmic_ambiguity}).
It also makes it difficult to ensure internal consistency of the estimations. Typically, this is solved through a DBN (see, \textit{e.g.} \cite{dbn}), which often assumes a constant time signature. DBN-free approaches are free of such constraints but can result in errors, such as duplicated or missing downbeats.

These shortcomings hardly surface in existing work: The F-score, often used as the primary metric of comparison, also treats beats and downbeats as independent events, and the datasets used for training and evaluation are dominated by pieces with constant time signature (97.7\% of training pieces for \cite{beatthis} have fewer than one time signature change per minute), most of which are 4/4 or 3/4 time (85\% and 9.6\% respectively for \cite{beatthis}), fitting the constraints of the DBN. AMLt and CMLt are reported to a lesser extent; they are missing from \cite{BeatNet} and not reported for the cross-validation in \cite{beatthis}. Furthermore, AMLt is designed for meters with groupings of two or four, again reflecting the bias in the data. This is mainly a problem in cases like 12/8 pieces that can be interpreted as 4/4 or vice versa, where the wrong guess is considered completely incorrect by AMLt as it does not consider groupings of three. 

The mathematically ideal approach would be to model the full distribution of all possible annotations of a piece, conditional on the audio. For a 30-second excerpt with 50 frames per second, each of which is either a downbeat, a non-downbeat beat, or no beat, this would amount to $3^{1500}$ possibilities, which is infeasible.
The previously described strategy makes this distribution tractable by factorizing it into $2\cdot1500$ framewise beat and downbeat probabilities, neglecting dependencies.
Here, we explore a solution in between: we assume that all relevant annotations can be factorized into a locally regular pulse, the \emph{tatum grid}, and a local tempo and time signature expressed in terms of that pulse, the \emph{subgrid} (a subset of the grid, not a subdivision). The term tatum is not strictly used in Bilmes \cite{tatum} definition as the smallest metric subdivision, but rather in accordance with the discussion provided by \cite{polak}, as an equal-pulse base, or a density referent that can be seen as underlying a metrical performance of music. To see the value of modelling a pulse that is potentially faster than the beat, consider the bridge in Here Comes the Sun by the Beatles. It features a time signature that can be interpreted as alternating between 4/4, 2/4, 3/8 and 5/8. This is well described by a constant eight-note tatum with beats that alternate between landing on every tatum and every second tatum. With the same tatum grid, other interpretations like alternating 4/4 and 5/4 are also possible, just by changing the subgrid.

In contrast to earlier hierarchical approaches \cite{seppanen2001,klapuri2006,jointbeattatum,durand2016,MultipleHypothesis}, which derive the pulse from the audio using hand-designed operations, we propose a differentiable approach to predict the local tatum grid and subgrid and integrate it with the Transformer of Beat This \cite{beatthis}.
Our model is designed to have a more expressive output than existing models, which could allow it to be used in more types of applications, for example when multiple time signature interpretations are desired. Despite this, we face similar limitations due to the biased data available for training and evaluation. Therefore, we do not expect it to surpass the best existing models on the standard test sets. Our primary objective is to compare the performance of our model directly with Beat This \cite{beatthis}, and document its advantages and shortcomings. 

We find that our model is comparable to Beat This for beats, but reaches significantly higher continuity metrics on downbeats, especially on pieces with changing time signatures. 
We believe that we are far from the performance ceiling of a grid-subgrid architecture. The estimated tatum-grid sometimes fails to include the true beats, and the final estimated metrical distributions are often inconsistent over time, particularly for pieces with time signature changes. Solving either of these problems should increase overall performance. We share our code and invite others to explore this.\footnote{\url{https://github.com/CPJKU/explore_this}} 

\section{Related work}\label{sec:related-work}

Multiple proposals follow the strategy of predicting framewise beat/downbeat probabilities and postprocessing those with Böck et al.'s DBN \cite{dbn}, successively using more powerful models:
While the original \cite{dbn} uses recurrent Long-Short Term Memory (LSTM) units,
Böck et al.\ \cite{deconstructanalyze} switched to Temporal Convolutional Networks,
Zhao et al.\ \cite{beattransformer} use Transformers with dilated self-attention on demixed spectrograms,
and Hung et al.\ \cite{spectnt} employ the Transformer-based SpecTNT architecture.
Beat This! \cite{beatthis} by Foscarin et al.\ also utilizes Transformers, but is trained with a Shift Tolerant Binary Cross Entropy (BCE) loss, which enabled state-of-the-art F-scores without a DBN.
A recent publication streak by Ru et al.\ investigates using pre-trained self-supervised Transformers without and with fine-tuning \cite{beatfm,hingenet}, sparsified self-attention \cite{BAT}, and Mamba \cite{beatmamba}, but sticks to the same basic strategy.

The idea of finding a pulse, then defining the beat or downbeat on top of it has already been used in early beat trackers such as \cite{seppanen2001,uhle2003,durand2016}.
A prominent technique for the first step is the Predominant Local Pulse (PLP) \cite{PLP}, which can be derived from a feature sequence through deterministic signal processing and is used for real-time beat tracking in \cite{realtimeplp}. The PLP can serve as a proxy for the tatum, but is not very robust when tempo or meter changes abruptly. Differentiable PLP \cite{dPLP} makes the PLP differentiable with respect to the underlying feature sequence, but is still a deterministic transformation.
In contrast, we model a distribution over the local tempo and phase of the tatum.

Gagneré et al.\ attempt to model ambiguity with a Multiple Hypothesis model \cite{MultipleHypothesis}. Like us, they create a grid and use it as a basis for guesses of different metrical structures, but their grid is based on the PLP. Pieces where the PLP fails, as with abrupt tempo or time signature jumps, are excluded from training. Furthermore, their model separately produces a small number of hypotheses each for beat and downbeat, while we model a joint distribution with a larger set of allowed time signatures. The biggest difference is that they only use multiple hypotheses for the self-supervised part of the training, favouring the common beat-activation strategy at the fine-tuning stage and during inference.

\section{Method}\label{sec:method}

\begin{figure}
    \centering
    \includegraphics[angle=90]{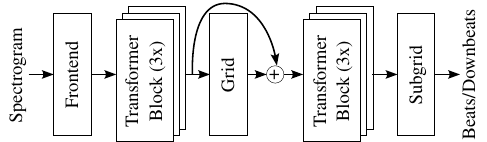}
    \caption{The overall architecture of the model. The frontend and Transformer blocks correspond to Beat This \cite{beatthis}.}
    \label{fig:summary}
\end{figure}

The architecture of our model is summarized in Figure \ref{fig:summary}. We use Beat This \cite{beatthis} as the basis, as it is open source and is among the highest-performing models not using a DBN. We insert a module that derives a tatum grid from the output of the first three Transformer blocks, and a module that predicts the subgrid after three more Transformer blocks.

\subsection{Pre-grid}
Before the tatum grid is estimated, the audio input (a 30-second mel spectrogram with 128 bins and 1500 frames at 50 frames per second) is processed by a frontend of alternating convolutions, frequency-wise and time-wise Transformer blocks \cite[Sec.~3.1.1]{beatthis} into a sequence of 1024-dimensional vectors. These are projected to $K$ dimensions for the main Transformer. Beat This sets $K\!=\!512$ for the large and $K\!=\!128$ for the small variant; we choose $K\!=\!256$ in between. We then pass the sequence through three of the six full Transformer blocks \cite[Sec.~3.1.2]{beatthis}.

\subsection{Grid}

\begin{figure}
    \centering
    
    \begin{subfigure}{\linewidth}
    \centering
    
    \includegraphics[width=0.9\linewidth]{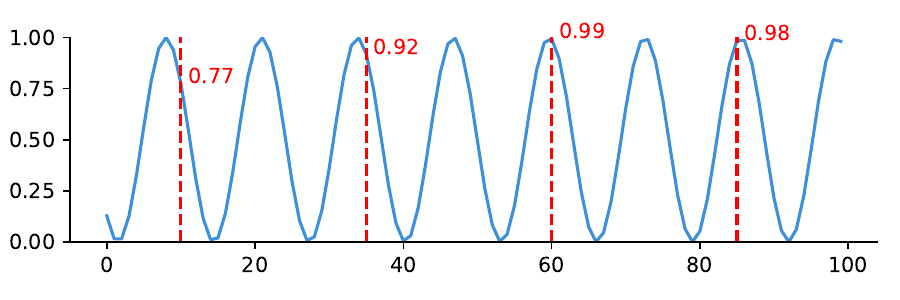}
    \caption{A wave with soft recall 0.92, close to a global maximum.}
    \end{subfigure}

    \begin{subfigure}{\linewidth}
    \centering
    
    \includegraphics[width=0.9\linewidth]{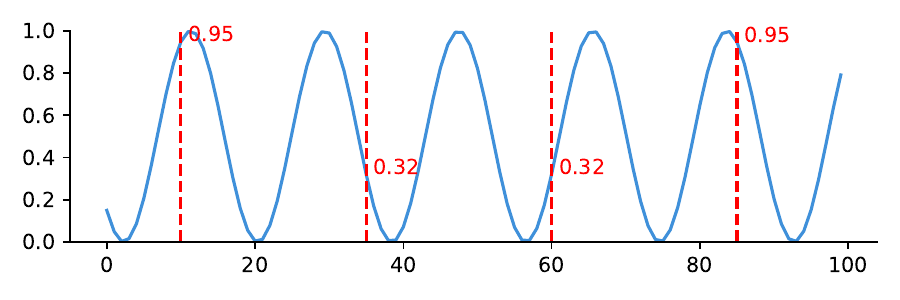}
    \caption{A wave with soft recall 0.62, stuck at a local maximum. }
    \end{subfigure}
    
    \begin{subfigure}{\linewidth}
    \centering
    
    \includegraphics[width=0.8\linewidth]{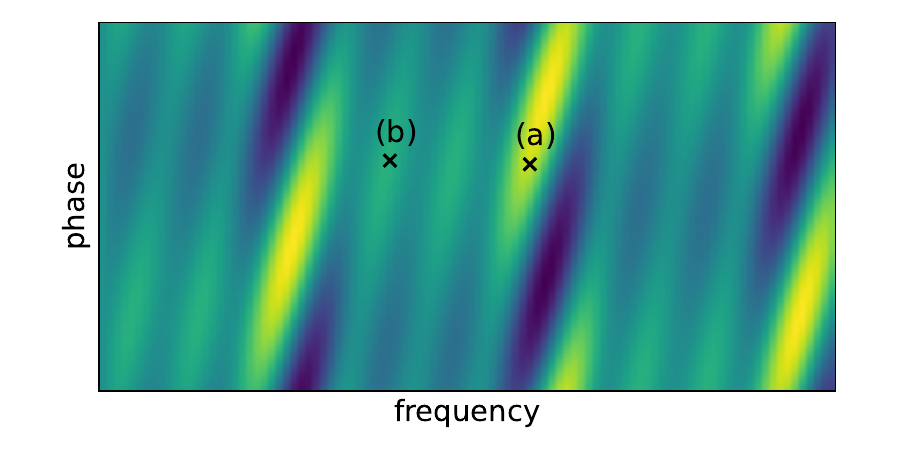}
    \caption{Soft Recall for a window with four ground truth beats, as a function of frequency and phase. There are bright maxima at integer multiples of the true frequency, where 100\% soft recall can be achieved. There are also local maxima, for example at (b). }
    \end{subfigure}
    \caption{Showcase of the properties of soft recall, applied to a single window. True beat positions are shown with red dashed lines in (a) and (b). Soft recall is differentiable with respect to phase and frequency of the predicted wave, but has local maxima with imperfect score as shown in (c), which means that soft-recall is non-concave. }
    \label{fig:soft_recall}
\end{figure}

From its $1500\!\times\!256$ input, the grid block tries to estimate the tatum grid, which is subject to several requirements:
(1) The grid should have 100\% recall on the beats, as any missed beats will be missed in the model's final prediction.
(2) 
The grid should not be faster than necessary,
to minimize the dimensionality of the ensuing subgrid prediction.
(3) The grid should reflect a consistent metrical level, like 8th or 16th notes, unless the underlying metric structure changes drastically.

To make (3) tractable, we choose the grid to be expressed as a tempo and phase, rather than independent framewise probabilities.
To allow for local fluctuations, we split the input into 1-second windows with individual predictions.
To enable a differentiable comparison against the beat annotations, for each window, we render a sine wave of the local tempo and phase, shifted to the range $[0,1]$.

\begin{figure}[t]
    \centering
    \includegraphics{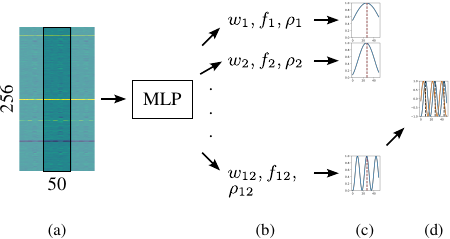}
    \caption{For a 1-second window of features (a), the grid block estimates weight, frequency and phase of 12 tatum candidates (b). Each defines a sine wave trained to maximize soft recall (c). The highest weighted one defines the tatum points and is passed on as a sine/cosine feature (d).}
    \label{fig:grid_demo}
\end{figure}
We then define the \textit{soft recall} as the mean of the sine wave evaluated at ground truth beats (Figure \ref{fig:soft_recall}), and set the training objective to maximize the soft recall. This objective is a differentiable way to achieve (1) and even (2), as the model will favour broader sine waves to combat noise in the annotations.

However, soft recall for a single wave is not concave. Depending on the number of true beats in a window, there can be a number of greater or smaller local maxima, as seen in Figure~\ref{fig:soft_recall}.
To help with the convexity of the loss, we use a mixture-of-experts model with multiple candidate waves (Figure~\ref{fig:grid_demo}).
The candidates span 12 non-overlapping tempo ranges, each covering a third of an octave, totalling in 4 octaves from 30--480 pulses per minute. Each candidate has one parameter to adjust the tempo inside the range, clamped by a sigmoid, and another to adjust the phase.
The distribution over candidates is modelled by a 12-way softmax.
Maximizing the weighted mean of the soft recalls for all candidates achieves convexity inside a single window, optimizing candidates and their weights.
For the backward pass, we mix in some uniformity to the weights, so each candidate receives a non-zero gradient to maximize soft recall. In summary, the model predicts twelve candidate tatums in non-overlapping frequency ranges, and a confidence distribution over them. 

To solve (3), we add a loss that penalizes inconsistencies in adjacent windows, while also being convex and differentiable. The inconsistency between two candidates in adjacent windows can be defined via the phase jump at the border and the difference in tempo. If the jump is sufficiently large, it stops mattering exactly how large, so we use soft clamping to reflect this. To define inconsistency between two full sets of 12 weighted candidates, we compute the pairwise inconsistency of all 66 pairs and then calculate the Earth Mover's Distance (EMD) between the distributions. To calculate EMD in a differentiable and batchable manner, we used the Sinkhorn algorithm \cite{sinkhornalgo} with a fixed number of iterations. 

In summary, the grid is described by a $W\!\times\!12\!\times\!3$ feature tensor, where $W\!=\!30$ is the number of windows (for our 30-second input and 1-second windows). For each window, the features encode a distribution over 12 possible candidate grids.
The 36 features per window are predicted from the corresponding $50\!\times\!256$ input excerpt (1/30th of the $1500\!\times\!256$ input of the grid block) through three linear layers with 50 hidden units of GELU activation.
The prediction loss is calculated using the weighted soft recall on the ground truth beats, and the regularization loss by estimating the inconsistencies between the predictions for adjacent windows. For the downstream use of the tatum grid, the most probable candidate is selected for each window. The windows are concatenated and smoothed with a Gaussian of 3 frames standard deviation to avoid double peaks at window borders. The local peaks of the resulting wave are the estimated tatum points, and the wave itself is passed on to predict the subgrid.
Specifically, the sine wave, the corresponding cosine wave and the mask of tatum points are projected to $256$ dimensions and added to the values of existing $1500\!\times\!256$ tensor.
After the remaining three transformer blocks of Beat This, we reach the subgrid block.

\subsection{Subgrid}
From the $1500\!\times\!256$ tensor that encodes both the audio and the soft tatum grid,
the subgrid block attempts to estimate the local metric distribution at every tatum point. We define a subgrid state as a set of four integers: Beat period $T_B$ represents how many tatums there are per beat, beat phase $\varphi_B \in\{1,...,T_B\}$ represents the index of the current tatum, while downbeat period $T_{D}$ and downbeat phase $\varphi_D \in\{1,...,T_D\}$ represent the period and index of downbeats in terms of beats. The offbeat after the second quarter note in a measure of 4/4 time could therefore be described as $(T_B, \varphi_B, T_D, \varphi_D)=(2,2,4,2)$ or $(4,3,4,2)$ depending on the choice of tatum. By setting the maximum $T_B\leq4$ and $T_D\leq7$ we limit the state space to 280 different possibilities. 

\begin{figure}[t]
    \centering
    \includegraphics[width=1\linewidth]{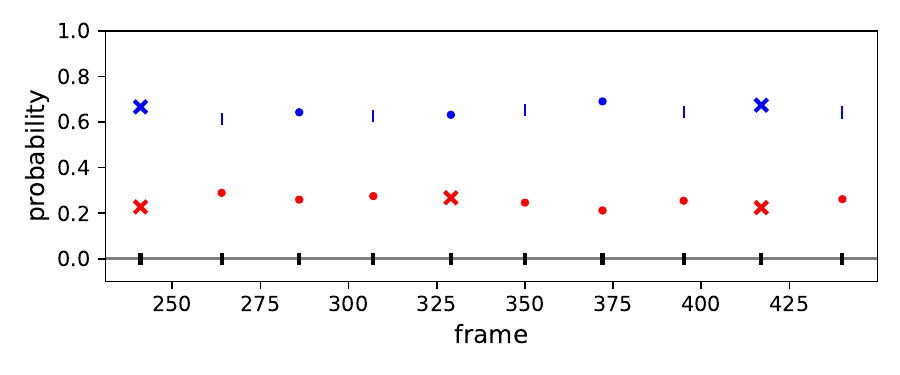}
    \caption{Visualization of the most probable subgrid states for the two most probable metric signatures. Shown in blue is [two tatums per beat, four beats per downbeat], and in red is [one tatum per beat, four beats per downbeat]. Only the most probable phase is shown in each case. Downbeats are marked with $\times$, beats with $\cdot$ and offbeats with $|$. The markings at $y=0$ denote the tatum grid.}
    \label{fig:subgrid_ambiguity}

\end{figure}

For every estimated tatum point, we predict a categorical distribution over these 280 bins (Figure \ref{fig:subgrid_bins}, Figure \ref{fig:subgrid_ambiguity}). The probability of a beat or downbeat can then be calculated by summing up all of the bins that imply a beat or downbeat, i.e., the ones with $\varphi_B=1$ and $\varphi_B=\varphi_D=1$, respectively. We calculate the prediction loss based on this probability and the ground truth annotations snapped to the nearest tatum point. Typically, this loss would be computed with Binary Cross Entropy, but due to unresolved numerical instabilities, we had to use Mean Squared Error. We need to snap the ground truth to the grid as the beat and downbeat probabilities are only meaningfully defined at the tatum points. If the tatum grid is good, this ensures that we can tolerate some noise in the exact timing of annotations, but on the negative side, the prediction loss will offer little guidance when the grid is bad.

As every subgrid state implies the state at every other tatum in that measure, we can directly measure the internal consistency of the subgrid by comparing the implied and estimated distributions at every tatum, and penalize the KL-divergence between the distributions. In order to allow the model to change time signatures, we treat all downbeats bins as equivalent for this loss. This means that a $(T_D, \varphi_D)=(4,4)$ that will usually be followed by a $(4,1)$ is also allowed to be followed by any other $(N, 1)$. 

\begin{figure}[t]
    \centering
    \begin{tikzpicture}[
    every node/.style={font=\small},
    box/.style={draw, minimum width=0.8cm, minimum height=0.8cm, text width=0.6cm, anchor=center},
    bigbox/.style={draw, thick}
]

\matrix (L) [matrix of nodes,
    nodes=box,
    column sep=-\pgflinewidth,
    row sep=-\pgflinewidth
] {
1\% &   -   &  -   &  -   \\
<1\% & 18\% &  -   &  -   \\
<1\% & <1\% & <1\% &  -   \\
<1\% & <1\% & <1\% & 80\% \\
};

\draw[thick] (L-1-1.north west) rectangle (L-4-4.south east);

\foreach \i in {1,...,4}
    \node[above=2pt] at (L-1-\i.north) {\i};

\foreach \i in {1,...,4}
    \node[left=2pt] at (L-\i-1.west) {\i};

\node[left=12pt, yshift=-10pt] at (L-2-1.west) {$T_D$};
\node[above=12pt, xshift=12pt] at (L-1-2.north) {$\varphi_D$};

\draw[red, thick] (L-4-4.north west) rectangle (L-4-4.south east);


\matrix (R) [matrix of nodes,
    nodes={box, minimum width=1.0cm, minimum height=1.0cm},
    column sep=-\pgflinewidth,
    row sep=-\pgflinewidth,
    right=0.5cm of L
] {
95\% &  -  \\
4\%  & 1\% \\
};

\draw[thick] (R-1-1.north west) rectangle (R-2-2.south east);

\foreach \i in {1,2}
    \node[above=2pt] at (R-1-\i.north) {\i};

\foreach \i in {1,2}
    \node[right=2pt] at (R-\i-2.east) {\i};

\node[right=12pt, yshift=-15pt] at (R-1-2.east) {$T_B$};
\node[above=12pt, xshift=15pt] at (R-1-1.north) {$\varphi_B$};


\draw[red, dashed, thick]
    (L-4-4.south east) -- (R-2-1.south west);

\draw[red, dashed, thick]
    (L-4-4.north east) -- (R-1-1.north west);
    
\end{tikzpicture}
    \vspace{-2em}
    \caption{A scaled-down example subgrid prediction for single grid point. Each output is a distribution over joint beat- and downbeat states, where a state represents the integer period and phase. For the true output, the left square is $7 \!\times\! 7$ and the right is $4 \!\times\! 4$.}
    \label{fig:subgrid_bins}
\end{figure}

\subsection{Inference}
Our model allows two ways to infer beats and downbeats for an audio excerpt.
Both begin by predicting the positions of the tatum points.
First, if we interpret the sequence of subgrid states at tatum points as a Markov model with fully known probabilities, we can compute the most probable total path with the Viterbi algorithm, with some uncertainty constant $s$ determining the probability of unexpected metric jumps. This will be referred to as Viterbi inference, and we will use $s=3/10$ in our experiments based on trial runs on the validation set.

Alternatively, we can calculate the total probability of a beat or downbeat at every tatum point as in the subgrid prediction loss, and use it to define a (very sparse) framewise activation function. We can then use simple thresholding or a DBN to materialize beats and downbeats. This neglects interdependences between successive tatum points, but is faster than Viterbi decoding. For simplicity, we apply a threshold of 0.5 on the framewise probabilities, and refer to this as threshold inference.

\section{Experiments}\label{sec:experiments}
We perform experiments to assess the quality of the tatum grid, to compare our model to the original Beat This for different subsets of the data, and to evaluate it against the state of the art.

\subsection{Training and evaluation}
For training and validation, we use the same datasets as Foscarin et al.\ \cite{beatthis}: ASAP~\cite{foscarin2020asap},
Ballroom~\cite{simac_ballroom,ballroom},
Beatles~\cite{beatles},
Candombe~\cite{candombe},
Filosax~\cite{filosax},
Groove MIDI~\cite{groove_midi},
GuitarSet~\cite{guitarset},
Hainsworth~\cite{hainsworth2004},
Harmonix~\cite{harmonix}, 
HJDB~\cite{hockman2012one},
JAAH~\cite{jaah},
Simac\cite{simac_ballroom},
RWC~\cite{rwc,rwc2} (classical, pop, royalty-free, and jazz),
SMC~\cite{smc},
TapCorrect~\cite{driedger2019towards}.
We use the exact same splits and training setup, following the public implementation. This includes using the exact same data augmentations and training hyperparameters, with 100 epochs and a batch size of 8.

For evaluation, we use standard beat tracking metrics F1, CMLt and AMLt~\cite{tempobeatdownbeat:book}, with default parameters as defined in the mir\_eval \cite{raffel2014mir_eval} library. We adopted the default behavior of mir\_eval, ignoring the first five seconds of every sample. To the best of our knowledge, this is also done by the other models we compare to. We also measure the recall and AMLt of our predicted grid. We primarily care about recall, as the recall is one of the primary targets and limits the theoretical maximum performance of the final predictions. AMLt is measured as an imperfect proxy for grid regularity. Though it falsely gives a bad score to tatum grids that are three or four times as fast as the beat, it can indicate grids that perfectly follow the true beats or twice the beat frequency. 

\subsection{Cross-validation}

\begin{table}[t!]
    \centering
    \small
    \caption{Grid metrics on cross validation dataset. Grid recall is measured as for the F1-score, by counting how many annotated beats have a grid point within 70\,ms. AMLt is calculated treating the grid points as beat predictions. } 
    \label{tab:grid_metrics}
    \setlength{\tabcolsep}{4pt}
\begin{tabular}{lcclcc}
    \toprule
      & Recall & AMLt & & Recall & AMLt \\
    \cmidrule(lr){1-3} \cmidrule(lr){4-6}
    ASAP & 80.9 & 56.3 
    &
    Harmonix & 96.9 & 92.6 
    \\
    Ballroom & 96.8 & 89.9 
    &
    HJDB & 95.8 & 96.6 
    \\
    Beatles & 97.3 & 89.0 
    &
    JAAH & 84.1 & 85.6 
    \\
    Candombe & 99.3 & 97.7 
    &
    RWC & 90.1 & 80.2 
    \\
    Filosax & 93.0 & 91.5 
    &
    Simac & 87.6 & 82.6 
    \\
    Gr. MIDI & 92.4 & 87.4 
    &
    SMC & 69.1 & 51.0 
    \\
    GuitarSet & 89.3 & 83.4 
    &
    TapCorrect & 96.6 & 85.9 
    \\
    Hainsworth & 93.3 & 90.9 
    &
    && \\
    \midrule
    All & 91.1 & 83.5 && \\
\end{tabular}

    \centering
    \small
    \caption{Results of 8-fold cross validation. We compare our model using threshold inference (Thr.), our model using Viterbi inference (Vit.) and the Beat This model (BT).
    } 
    \label{tab:crossval}
    \begin{tabular}{lcccccc}
    \toprule
    & \multicolumn{3}{c}{Beat CMLt} & \multicolumn{3}{c}{Downbeat CMLt} \\
    \cmidrule(lr){2-4} \cmidrule(lr){5-7}
    & Thr. & Vit. & \textit{BT} & Thr. & Vit. & \textit{BT} \\
    
    \midrule
    ASAP            & \textbf{50.4}  &   50.1   & 50.3          &  \textbf{31.2}  &  28.6           &  23.5       \\
    Ballroom        & 92.8           &   90.6   & \textbf{96.4} &  \textbf{95.0}  &  92.5           &  92.9         \\
    Beatles         & 84.8           &   83.3   & \textbf{87.2} &  \textbf{77.8}  &  76.1           &  73.7         \\
    Candombe        & 98.9           &   98.3   & \textbf{99.8} &  99.6   &  99.3          &  \textbf{99.9}    \\
    Filosax         & 85.5           &   84.9   & \textbf{98.8} &  \textbf{98.7}  &  97.7           &  96.1        \\
    Gr. MIDI        & 82.9           &   83.4   & \textbf{84.9} &  \textbf{77.5}  &  76.8              &  70.9    \\
    GuitarSet       & \textbf{85.0}  &   84.5   & 82.4          &  \textbf{89.6}  &  86.9           &  79.8         \\
    Hainsworth      & 81.2           &   81.4   & \textbf{84.0} &  73.4           &  \textbf{73.6}  &  63.6          \\
    Harmonix        & 88.5          &   87.9   & \textbf{89.9} & \textbf{85.2}  &  84.5           &  81.2           \\
    HJDB            & 94.6           &   94.6   & \textbf{97.2} &  94.2           &  94.3           &  \textbf{94.8}  \\
    JAAH            & 68.5           &   68.1   & \textbf{88.5} &  \textbf{78.8}  &  78.3           &  71.5           \\
    RWC             & 74.3          &   73.3   & \textbf{75.7} &  \textbf{72.3}  &  70.6           &  69.0           \\
    Simac           & \textbf{72.7}  &   72.6   & 55.8          &   -             &   -             &  -            \\
    SMC             & 48.0           &   48.5   & \textbf{51.4} &   -             &   -             &  -              \\
    TapCorrect      & \textbf{82.2}  &   79.8   & 81.9          &  \textbf{79.3}  &  74.8           &  68.2          \\
    \midrule
    All         & \textbf{79.2}  &  78.5    & 79.0          & \textbf{78.4}   & 76.9            & 73.5           
\end{tabular}

    \centering
    \small
    \caption{Results on different subsets of cross validation data, excluding datasets without downbeat annotations. Pure refers to samples with constant number of beats per downbeat. 4/4 refers to samples with 4 beats per downbeat. Changing refers to samples with more than two bar length changes per minute. }
    \label{tab:time_signature_res}
    \begin{tabular}{lccc}
    \toprule
      & Pure 4/4 & Pure non-4/4 & Changing \\
    \midrule
    Share of data & 75.7\% & 13.3\% & 1.6\% \\
    No.\ of pieces & 2836 & 499 & 59 \\
    \midrule
    & \multicolumn{3}{c}{Downbeat CMLt} \\
    \cmidrule(lr){2-4}
    Threshold & \textbf{84.0} & \textbf{56.0} & 46.9 \\
    Viterbi & 82.6 & 54.3 & \textbf{47.3} \\
    Beat This \cite{beatthis} & 78.9 & 54.4 & 36.4 \\
    \bottomrule
\end{tabular}

\end{table}

First, we do an 8-fold cross-validation on the training data to 
measure the performance of the tatum grid, as shown in Table \ref{tab:grid_metrics}. For the ASAP and SMC datasets, we see that the recall is good but the AMLt is bad. This suggests that the tatum grid is very fast or not regular in terms of beats. For the HJDB and JAAH datasets, we have the opposite: AMLt is higher than recall. This happens when the tatum grid is half the frequency of the beats.

Qualitative analysis of the grid predictions showed that the grid struggled with tempo dynamics that are common in classical music. This is unavoidable with the architecture we have chosen, as we only allow the tatum to change tempo once a second. We also do not directly penalize irregularity in the number of tatums per beat, so abrupt slowdowns in tempo are often handled by the tatum speeding up, catching the next beat after generating an incorrect extra tatum.

\begin{table*}
    \centering
    \caption{Results on GTZAN for the final model, compared with state-of-the-art models with and without DBN postprocessing. The best score in each metric is marked in \textbf{bold} for each of the two categories.}
    \label{tab:final_res}
    \begin{tabular}{lcccccc}
    \toprule
    & \multicolumn{3}{c}{Beat} & \multicolumn{3}{c}{Downbeat}  \\
    \cmidrule(lr){2-4} \cmidrule(lr){5-7}
                    & F1       & CMLt            & AMLt            & F1      & CMLt           & AMLt \\
    \midrule
    \textit{DBN-free} &&&&&& \\
    Our (Threshold)       & 87.9 $\pm$ 0.8 & 79.8 $\pm$ 1.2 & 90.3 $\pm$ 0.1  
                    & 77.3 $\pm$ 0.9 & 72.4 $\pm$ 0.7 & 85.1 $\pm$ 0.6 \\
    
    Our (Viterbi)         & 88.0 $\pm$ 0.7 & 80.0 $\pm$ 1.1 & 90.4 $\pm$ 0.1  
                    & 77.4 $\pm$ 1.0 & \textbf{72.7} $\pm$ 0.8 & \textbf{86.2} $\pm$ 0.4 \\

    Beat This \cite{beatthis}      & 89.1 $\pm$ 0.3 & 79.8 $\pm$ 0.6 & 89.8 $\pm$ 0.4  
                    & \textbf{78.3} $\pm$ 0.4 & 67.3 $\pm$ 0.8 & 79.1 $\pm$ 0.6 
                     \\
                    
    Gagneré et al. \cite{MultipleHypothesis}  ST-BCE & \textbf{89.6} & \textbf{81.8} & \textbf{91.8}  
                    & 77.5 & 67.0 & 80.8 
                     \\
    \midrule
    \textit{With DBN} &&&&&& \\
    
    Beat This + DBN & 88.1 $\pm$ 0.3 & 80.5 $\pm$ 0.4 & 91.1 $\pm$ 0.2 
                    & 77.4 $\pm$ 0.2 & 73.3 $\pm$ 0.2 & 87.8 $\pm$ 0.5 \\
    
    Hung et al. \cite{spectnt}     & 88.7 & 81.2 & 92.0  
                    & 75.6 & 71.5 & 88.1 
                     \\
    Gagneré et al. BCE-DBN & \textbf{89.6} & \textbf{82.6} & \textbf{92.5} & \textbf{78.3} & \textbf{74.7} & \textbf{88.2} \\

    \bottomrule
\end{tabular}

\end{table*}

Using the same cross-validation models, we evaluate their beat and downbeat estimations in terms of CMLt, shown in Table~\ref{tab:crossval}.
We compare the threshold and Viterbi inference algorithms with the cross-validation results from Beat This (which were calculated using their open-source models corresponding to each fold). In general, our models perform on par with or slightly below Beat This on beats, but perform much better on downbeats. Threshold inference gives a slightly better score than Viterbi in most cases.

To show the effect of the class imbalance between time signatures in beat tracking datasets, we also split the cross validation pieces into categories based on time signatures and show the corresponding results in Table~\ref{tab:time_signature_res}. Performance plummets for all models when evaluating on music that is not in 4/4 time, but ours seems a bit more stable for changing time signatures.

\subsection{Final test}

For the final test, we train three models with different seeds on all of the training data, testing on the held-out test dataset GTZAN \cite{gtzan}.
The metrics for other models were taken from their respective papers.
The final results are shown in Table \ref{tab:final_res}. We compare both of our inference methods (thresholding and Viterbi) with state-of-the-art models with and without DBN postprocessing. Our model performs worse than the best DBN models across the board, but beats other DBN-free models on downbeat continuity metrics by a significant margin. The Viterbi inference is slightly better than threshold across the board, with the greatest improvements displayed for downbeat continuity metrics. Not shown in the table are our grid recall: $90.5 \pm 2.0$ and grid AMLt: $89.6 \pm 0.6$.

\section{Conclusion}\label{sec:discussion}

Using our algorithm, we provide a proof-of-concept for a novel approach to model the problem of beat tracking. During development, many hurdles had to be overcome, and at the time of writing we are convinced that almost every part of the architecture could be improved. For example, having grid-windows shorter than 1 second would be required to better capture timing fluctuations, but currently causes problems with loss convexity. Another problem is the unresolved numerical issues with BCE that forced us to use MSE in the subgrid. The presented model is also very sensitive to the choice of hyperparameters for the grid and subgrid modules. However, we strongly believe that these problems are not inherent to the strategy, and can be solved with further development. 

Surprisingly, our model performs about as well with threshold inference as when using Viterbi. The threshold inference is a variant of a framewise probability approach, which we hypothesized to have significant limitations. The inability of the Viterbi interpretation to perform better implies that either the datasets have too few ambiguous samples to make a difference, or that something in our design prevents the model from obtaining its full expressive power. The fact that the threshold inference managed to beat Foscarin et al.\ \cite{beatthis} on continuity metrics nonetheless shows the regularizing power of theory-informed design. 

One advantage of our problem formulation is that it is highly separable. Focused efforts to improve the learning of the tatum grid could be done independently of improvements to the subgrid. There is also potential for the learned tatum to replace PLP in subgrid-like strategies such as \cite{MultipleHypothesis}. 

Explicitly modelling a distribution is not the only way to encode ambiguity using Deep Learning. Generative models can also be used to sample from an implicit distribution. This path could be a way to avoid a lot of the struggles associated with our method, but comes with its own challenges. 
Nonetheless, we believe that the era of independent framewise beat and downbeat probabilities will have to end at some point, as they cannot encode ambiguity, and therefore cannot capture the richness of metrical perception.

\section{Acknowledgments}

This research was funded in part by the Austrian Science Fund (FWF) [10.55776/P36653], by the Swedish Research Council under Grant 2024-01832, and by the Wallenberg AI, Autonomous Systems and Software Program—Humanities and Society (WASP-HS) funded by the Marianne and Marcus Wallenberg Foundation.

\bibliography{ISMIRtemplate}

\end{document}